%% file: DipoleMany.tex
\documentclass[aip,jcp,reprint]{revtex4-2}

\usepackage{graphicx}
\usepackage[utf8]{inputenc}
\usepackage{amsmath}
\usepackage{verbatim}
\usepackage{amssymb}
\usepackage{listings}
\usepackage{color}
\usepackage{hyperref}
\usepackage{amsmath}
\usepackage{mathtools}
\usepackage{color}
\usepackage{xspace}
\usepackage{siunitx}
\newcommand{\pit}[1]{\pi^{\{#1\}}}

\newcommand{\dens}{D}
\newcommand{\tint}{\tau_{\text{int}}}
\newcommand{\tchain}{\tau_{\text{chain}}}
\newcommand{\pfinal}{\pvec(t)}
\newcommand{\pinitial}{\pvec(\tresamp)}
\newcommand{\pinitialp}{\pvec'(\tresamp)}
\newcommand{\tresamp}{t_{\text{res}}}

\newcommand\subfig[2]{{Fig.~\ref{#1}{#2}}}
\usepackage{bold-extra}

\input{latexcommands_2021d.tex}

\begin{document}

\title[]{Hard-disk dipoles and non-reversible Markov chains}

\author{Philipp~Höllmer}
\email{hoellmer@physik.uni-bonn.de}
\affiliation{Bethe Center for Theoretical Physics, University of Bonn, 
Nussallee 12, 53115 Bonn, Germany}

\author{A.~C.~Maggs}
\email{anthony.maggs@espci.fr}
\affiliation{CNRS UMR7083, ESPCI Paris, Université PSL, 10 rue 
Vauquelin, 75005 Paris, France}

\author{Werner~Krauth}
\email{werner.krauth@ens.fr}
\affiliation{Laboratoire de Physique de l’Ecole normale supérieure, ENS, 
Université PSL, CNRS, Sorbonne Université, Université de Paris, Paris, France}

\date{\today}
	
\begin{abstract}
We benchmark event-chain Monte Carlo (ECMC) algorithms for tethered hard-disk 
dipoles in two dimensions in view of application of ECMC to  water 
models in molecular simulation. We characterize the rotation dynamics of 
dipoles through the integrated autocorrelation times of the polarization. The
non-reversible straight, reflective, forward, and Newtonian ECMC algorithms
are all event-driven, and they differ only 
in their update rules at event times. They realize considerable speedups with 
respect to the local reversible Metropolis algorithm. We also find 
significant speed differences among the ECMC variants.
Newtonian ECMC appears particularly well-suited for overcoming 
the dynamical arrest that has plagued straight ECMC for three-dimensional 
dipolar models with Coulomb interactions. 
\end{abstract}

\maketitle

\section{Introduction}

Markov-chain Monte Carlo~\cite{Metropolis1953} (MCMC) is a
computational method for 
sampling high-dimensional probability distributions $\pi$ including the 
Boltzmann distribution of statistical physics. 
In MCMC, an initial sample $x$, at a time $t=0$, is drawn from a distribution 
$\pit{0}$ rather than from the \quot{equilibrium} distribution $\pi$. The sample 
$x$, at time $t$, moves to the sample $y$, at time $t+1$, with a 
conditional probability $P_{xy}$, element of a time-independent transition 
matrix $P$. The probability distribution  $\pit{t}$ then moves to
$\pit{t+1} = \pit{t} P$. The intermediate probability distributions are usually 
not known explicitly, but they evolve towards $\pi$ for $t \to \infty$. The 
transition matrix
$P$ must satisfy a balance condition that expresses the stationarity of 
$\pi$. Reversible transition matrices satisfy the detailed-balance condition 
$\pi_x P_{xy} = \pi_y P_{yx}$ for all $x$ and $y$. Detailed balance is 
equivalent to the statement that the equilibrium flow $\FCAL_{xy} = \pi_x 
P_{xy}$ from $x$ to $y$ equals the reverse flow from $y$ to $x$. Non-reversible 
MCMC 
algorithms satisfy a weaker global-balance condition $\pi_x = \sum_y \pi_y 
P_{yx}$ for all $x$. Under conditions of irreducibility and 
aperiodicity,~\cite{Levin2008} they converge to the equilibrium 
distribution $\pi$ with non-zero net flows $\FCAL_{xy}- \FCAL_{yx}$. 
Samples
are then drawn from the equilibrium distribution $\pi$ for $t \to \infty$ by 
a non-equilibrium random process with non-vanishing net flows. 

In past decades, research and applications have focused almost exclusively on 
reversible Markov chains (and some close relatives, such as sequential 
schemes~\cite{Berg2004book,RenOrkoulas2007}). Reversible Markov chains are 
straightforward to set up for arbitrary probability distributions $\pi$, and are 
easy to conceptualize, in particular because of the real-valued eigenvalue 
spectrum of their transition matrices. Popular reversible Markov chains include 
the Metropolis-~\cite{Metropolis1953} or 
heat-bath~\cite{Glauber1963,Creutz1980HeatBath,Geman1984} algorithms. Interest 
in non-reversible Markov chains has increased in recent years, ever since it was 
understood~\cite{Diaconis2000,Chen1999} that such algorithms often approach 
equilibrium faster than  reversible Markov chains. Systematic non-reversible 
Markov-chain schemes are becoming available.~\cite{Turitsyn2011,Bernard2009}

Event-chain Monte Carlo~\cite{Bernard2009,Michel2014JCP,Krauth2021eventchain} 
(ECMC) is a family of non-reversible local Markov chains that have been applied 
to, e.g., particle~\cite{Bernard2011} and spin 
systems,~\cite{MichelMayerKrauth2015,Nishikawa2015,KimuraHiguchi2017} 
polymers~\cite{Mueller2020,Kampmann2021} and field-theoretical 
models.~\cite{PhysRevD.98.054502} ECMC algorithms are defined in continuous 
time, and they can be interpreted as piecewise deterministic Markov 
processes.~\cite{Davis1984} They have found applications in Bayesian 
statistics.~\cite{Bierkens2019, BouchardCote2018} A number of features set ECMC 
algorithms apart from reversible Markov chains. First, they may reach the 
\quot{equilibrium} distribution $\pi$ on fast ballistic time scales rather than 
on the diffusive time scales that are typically associated with local 
reversibility. Considerable speedups were demonstrated in analytically solvable 
test cases~\cite{Diaconis2000,Chen1999,Lei2018_OneD} and recovered in 
applications of practical 
relevance.~\cite{Bernard2011,Kampmann2021,Krauth2021eventchain, 
PhysRevD.98.054502} Second, ECMC,  for a given model (probability distribution 
$\pi$), covers closely related local MCMC algorithms that can behave quite 
differently.
Algorithms adopt a wide range of factor sets for the 
breakup of the Boltzmann distribution. On a finer scale, even for a given set of 
factors, different variants may have widely different behaviors, as we will 
discuss in the present paper. Third, ECMC need not evaluate the ratio $\pi_y / 
\pi_x$ to move from a sample $x$ to a sample $y$ (in other words, need not 
evaluate the  change in energy), because the factors are  statistically 
independent.~\cite{Michel2014JCP,Faulkner2018} 

The sampling of the Boltzmann distribution without evaluating the energy is 
implemented in a general-purpose open-source ECMC 
application~\cite{Hoellmer2020} for classical molecular simulation in Coulomb 
systems, where the evaluation of the energy represents for other 
methods the 
computational bottleneck. In the test case of three-dimensional water with the 
simple-point-charge-flexible-water (SPC/Fw) potential~\cite{WuTepperVoth2006} 
and the \quot{periodic} variant of straight ECMC (see \sect{subsec:Straight} for 
a definition), the simulated water molecules however resisted rotation and the 
polarization remained dynamically arrested for long times.~\cite{Qin2020Thesis} 
The present paper studies many variants of ECMC for  a simpler model of polar 
molecules, and suggests that its periodic variant is ill-suited to 
these systems. 

In a generic Markov chain, the sample space $\Omega$ is distinct from the moves 
$\Delta$. The latter are then part of a move set and  lead from  
samples 
$x \in \Omega$  to  samples $y\in \Omega$. In the local reversible Metropolis 
algorithm for particle systems, for example, $\Delta = (\delta, i)$ consists in 
a small random displacement $\delta$ of a random particle 
$i$. 
ECMC, in contrast, belongs to the class of  lifted Markov 
chains,~\cite{Diaconis2000, Chen1999} in which parts of the move set
are integrated into a \quot{lifted} sample space. In the above example, the 
proposed  move $\Delta'= (\delta', i')$  at time $t+1$, after a 
move $\Delta$, at time $t$, is then no longer independently sampled, but it 
rather depends on the lifted configuration, and in particular on $\Delta$. The 
probability distribution lives itself in an extended space. Nevertheless, in 
the cases that we consider here, the stationary probability 
distribution 
in the lifted space factorizes into the 
original distribution $\pi$ and a distribution for the lifting variables, while 
the transition matrix non-trivially couples the two sectors. 

In this paper we  caricature the SPC/Fw water model as tethered hard-disk 
dipoles. This replaces a three-dimensional model by a two-dimensional one, and 
lumps the vibrational, bending, Lennard-Jones, and Coulomb interactions into a 
hard-disk potential. The simplified dipole model  retains the polar nature of 
molecules, and it can be simulated orders of magnitude faster than the system 
of 
flexible Coulomb dipoles. It also greatly simplifies the implementation of the  
different algorithms. 
This allows us to  scan straight,~\cite{Bernard2009, Qin2020fastsequential} 
reflective,~\cite{Bernard2009} forward,~\cite{Michel2020} and 
Newtonian~\cite{Klement2019} ECMC with thousands of  parameter sets. We 
benchmark the ECMC algorithms against the  local Metropolis algorithms with 
different displacement sets. We characterize the speed of algorithms \emph{via} 
the autocorrelation of the polarization, the mode that with ECMC equilibrates 
slowly in the SPC/Fw water model.~\cite{Qin2020Thesis}

For dipole-model parameters that roughly correspond to those of SPC/Fw water, we 
find a $50$-fold speedup for the fastest non-reversible ECMC algorithm 
with respect to the optimized local reversible Metropolis algorithm. We also 
find a more than order-of-magnitude speed difference among the different ECMC 
variants. The different versions of straight ECMC are clearly the slowest 
ones
at low density, while they are comparable with  forward and reflective ECMC at 
high densities. Newtonian ECMC is by far the fastest. We suggest 
that this is due to its absence of intrinsic parameters. 

The content of this paper is as follows: In \sect{sec:DipoleModelDefinition},  
we define the two-dimensional tethered hard-disk dipole model and motivate its 
parameters with respect to physical systems of the SPC/Fw model in three 
dimensions. We also discuss the polarization, which tracks the ability of 
dipoles to rotate. In \sect{sec:ECMCDescription}, we discuss the
reversible and  non-reversible MCMC  algorithms that we have implemented and 
illustrate their behavior 
for the case of a single dipole. We also discuss subtle aspects concerning the 
irreducibility of the Markov chains. In 
\sect{sec:Autocorrelations}, we benchmark the different variants of 
ECMC against the local reversible Metropolis algorithm for a range of densities 
and system sizes, and we evidence the superiority of the non-reversible ECMC 
algorithms.
In \sect{sec:Conclusions}, we project how our conclusions can be extended to the 
more complex water models.

\section{Hard-disk dipole model}
\label{sec:DipoleModelDefinition}

\begin{figure}
	\centering
	\includegraphics{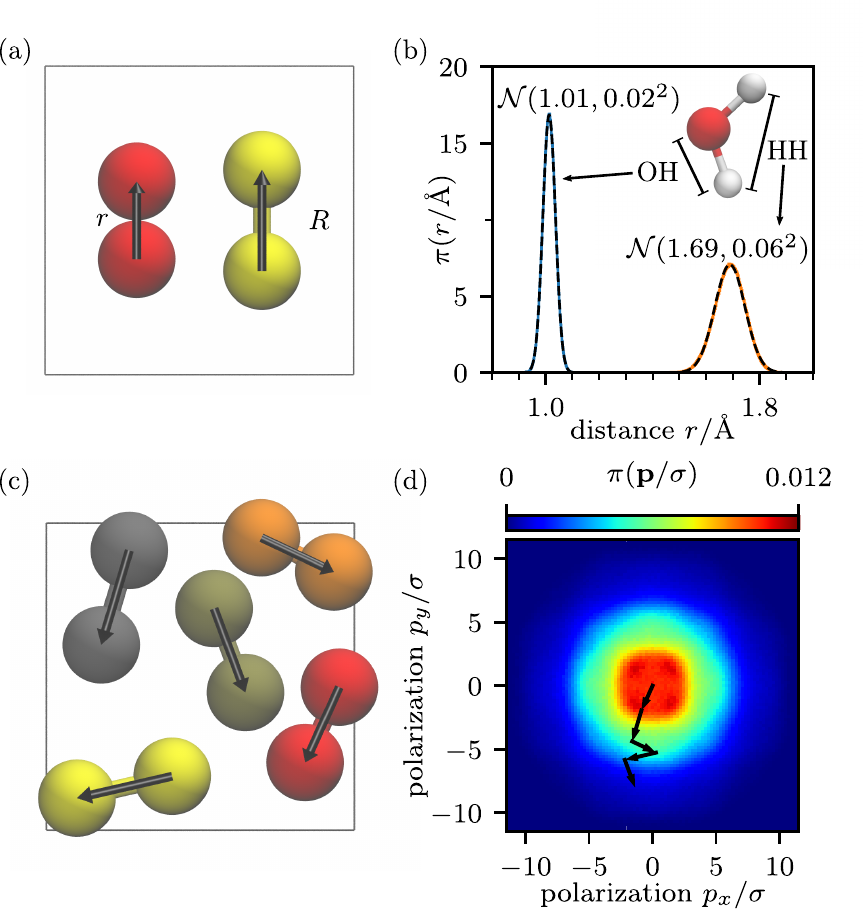}
	\caption{Tethered hard-disk dipole model. 
		\subcap{a} Dipoles with two hard disks of radius $\sigma$, and 
		minimum extension $r = 2 \sigma$ and tether length 
		$R$. 
		\subcap{b} Probability distribution of the oxygen--hydrogen (OH) 
		and hydrogen--hydrogen (HH) bond lengths for a single SPC/Fw water 
		molecule at room temperature.
		\subcap{c} Five dipoles with polarization  $\pvec_i$. 
		\subcap{d} Total polarization $\pvec$ of (c), 
		together with its probability distribution.}
	\label{fig:DefinitionDipole}
\end{figure}

We consider tethered dipoles in a two-dimensional square box of sides $L$ with 
periodic boundary conditions. Each dipole consists of two hard disks of radius 
$\sigma$, with a flat inner-dipole interaction that 
constrains its extension $\rho$ (the separation of the disk centers) between 
the 
contact distance  $r  = 2 \sigma$ and the tether length   $ R$ (see 
\subfig{fig:DefinitionDipole}{a}). 
Dipole configurations without overlapping disks and with all dipole extensions 
between $r$ 
and $R$ all have the same Boltzmann weight, whereas all other 
configurations have zero statistical weight. As the minimum  dipole extension
at contact equals $2\sigma$ (see 
\subfig{fig:DefinitionDipole}{a}), each dipole configuration is also a 
configuration of hard disks. Any dipole system is parameterized by the number of 
dipoles $N$, the tether ratio $\eta =  R/r$, and the hard-disk 
density $\dens = 2N\pi\sigma^2/L^2$. 

In a single three-dimensional water molecule at room temperature 
(as for example described in the SPC/Fw model), the 
oxygen--hydrogen distance fluctuates by $\sim \SI{2.3}{\percent}$ and the 
hydrogen--hydrogen distance by $\sim \SI{3.4}{\percent}$ (see 
\subfig{fig:DefinitionDipole}{b}). In the hard-disk dipole model, we thus adopt 
a tether ratio $\eta=1.1$, for which the $\sim \SI{2.75}{\percent}$ fluctuations 
of the extension are quite similar.  The tether length $R$ is thus only 
$10\%$ larger than the contact distance between two  disks, so that two dipoles 
cannot lock into a crossed state that would be  difficult to disentangle in two 
dimensions. Our results for the dynamics of two-dimensional models may well
extend to three dimensions. 

We measure dynamical information through the integrated autocorrelation 
functions of components of the total polarization $\pvec$:
\begin{equation}
 \pvec = \sum_{i = 1}^{N} \pvec_i,
\label{equ:PolarizationDefinition} 
\end{equation}
where $\pvec_i$ is the oriented separation vector between the two disks in 
the 
$i$th dipole, possibly corrected for periodic boundary conditions (see 
\subfig{fig:DefinitionDipole}{c and d}). 

For $N=1$, analytic results are available for the dynamics of some of the 
algorithms discussed here.~\cite{Qin2020fastsequential} The center-of-mass 
motion of the single dipole then factors out, and the sample space $\Omega$ 
simply corresponds to the possible values of the polarization vector, in other 
words to the two-dimensional ring with inner radius $r$ and outer radius $R$ 
(see Ref.~\onlinecite{Qin2020fastsequential}). The single-dipole polarization 
$\pvec$ is uniformly distributed  in this ring.

\section{Reversible and non-reversible MCMC}
\label{sec:ECMCDescription}

In the present section, we introduce MCMC algorithms that will be used  in 
\sect{sec:Autocorrelations}. We also illustrate them for a single dipole 
($N=1$) and review some of their fundamental properties. We only consider 
algorithms where at each moment a single disk moves, as this can also be 
realized within ECMC in the presence of long-range Coulomb 
interactions.~\cite{Faulkner2018,Hoellmer2020} Dipole rotations---that relax the 
polarization---are thus pieced together from subsequent displacements of single 
disks. This has already proven efficient in ECMC for polymer models and for 
elongated hard needles.~\cite{Mueller2020, Kampmann2021} Explicit dipole 
rotations appear difficult to put into place in this context. 

The local MCMC moves discussed here feature small 
single-disk displacements at each time step (for the Metropolis algorithm) or 
continuous moves in time (for the ECMC algorithms). Again, this is 
motivated by the future applications to more complicated systems such as water, 
for which non-local moves including cluster 
updates~\cite{Dress1995,LiuLuijten2004} appear  out of reach. Although we thus 
restrict our attention to a subclass of MCMC algorithms with very similar 
design principles, we will show in \sect{sec:Autocorrelations} that their 
properties differ substantially.

In the ECMC algorithms, a single active disk moves with a constant velocity 
in a given direction. The straight-line trajectories stop at 
events which correspond to hard-disk collisions or to the moment when the 
dipole extension 
reaches the tether length. In addition, resamplings take place at predefined 
MCMC times, which are separated by a chain time $\tchain$. Resamplings also 
interrupt the straight-line trajectory, and for example draw a new velocity 
and a new active disk. 

\subsection{Reversible local Metropolis algorithms}

The local Metropolis algorithm,~\cite{Metropolis1953}  for the dipole model, 
selects a random disk $i$ at each time step. For the  cross-shaped displacement 
set, the proposed move for $i$ is sampled uniformly between $-\delta$ and 
$\delta$ with range $\delta$, randomly in $x$- or in $y$-direction. For the 
square-shaped displacement set, the proposed move is uniformly  sampled from a 
square of sides $2 \delta$ centered at zero, with both the $x$ and the $y$ 
components of the displacement sampled independently between $-\delta$ and 
$\delta$. In both cases, the move is accepted if it violates no constraints. If 
the move is rejected, the configurations at times $t$ and $t+1$ are the same. 
The Metropolis algorithm, with the given displacement sets, satisfies the 
detailed-balance condition. For $N=1$, the polarization $\pvec$ performs a 
random walk in the sample space $\Omega$. The algorithm can be considered 
irreducible for all $N$, though some blocked configurations~\cite{Boroczky1964} 
may pose problems.~\cite{Hoellmer2021}

\subsection{Straight ECMC}
\label{subsec:Straight}

\begin{figure}
    \centering
    \includegraphics{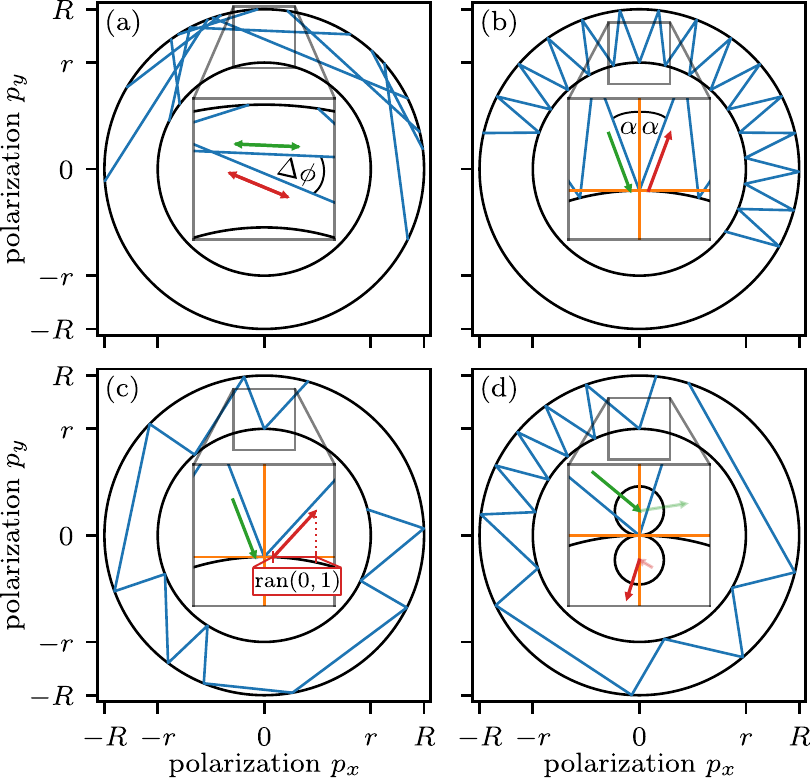} \caption{Sample trajectories 
for a single tethered dipole. The polarization $\pvec$ is the oriented 
vector between the two disk centers. \subcap{a} Sequential variant of 
straight  ECMC.~\cite{Qin2020fastsequential} \subcap{b} Reflective 
ECMC.~\cite{Bernard2009} \subcap{c} Forward ECMC.~\cite{Michel2020} \subcap{d} 
Newtonian ECMC.~\cite{Klement2019}
}
\label{fig:trajectories}
\end{figure}

In straight ECMC,~\cite{Bernard2009} the velocity is of constant absolute 
value, 
and it changes only at resamplings. We analyze several variants for updating the 
angle $\phi$ that describes the velocity in two dimensions. In the 
\quot{periodic} variant, the angle alternates between $\phi = 0 $ and $\phi = 
\pi/2$. In arbitrary dimension, the velocity cycles through the unit vectors 
aligned with the positive coordinate axes. The periodic variant, used in most 
previous works, is implemented in the general-purpose ECMC 
application.~\cite{Hoellmer2020} In the present paper, we also analyze the 
\quot{random} variant where $\phi$ is uniformly sampled in the interval $[0, 
2\pi)$ and, finally, the \quot{sequential} variant,\cite{Qin2020fastsequential} 
where $\phi$ is incremented by a constant $\Delta \phi$ at each resampling. The 
polarization trajectory between resamplings of straight ECMC is a straight line. 
In the periodic variant, subsequent straight-line trajectories are at right 
angles. They are at an angle $\Delta\phi$ in the sequential variant (see 
\subfig{fig:trajectories}{a} for an example in $N=1$). The described variants 
have very different time evolutions. The polarization trajectories of the 
sequential variant, for example,  persistently rotate for $N=1$,  with a total 
rotation angle that diverges as $\Delta \phi$ goes to 
zero.~\cite{Qin2020fastsequential}

\begin{figure}
    \centering
    \includegraphics{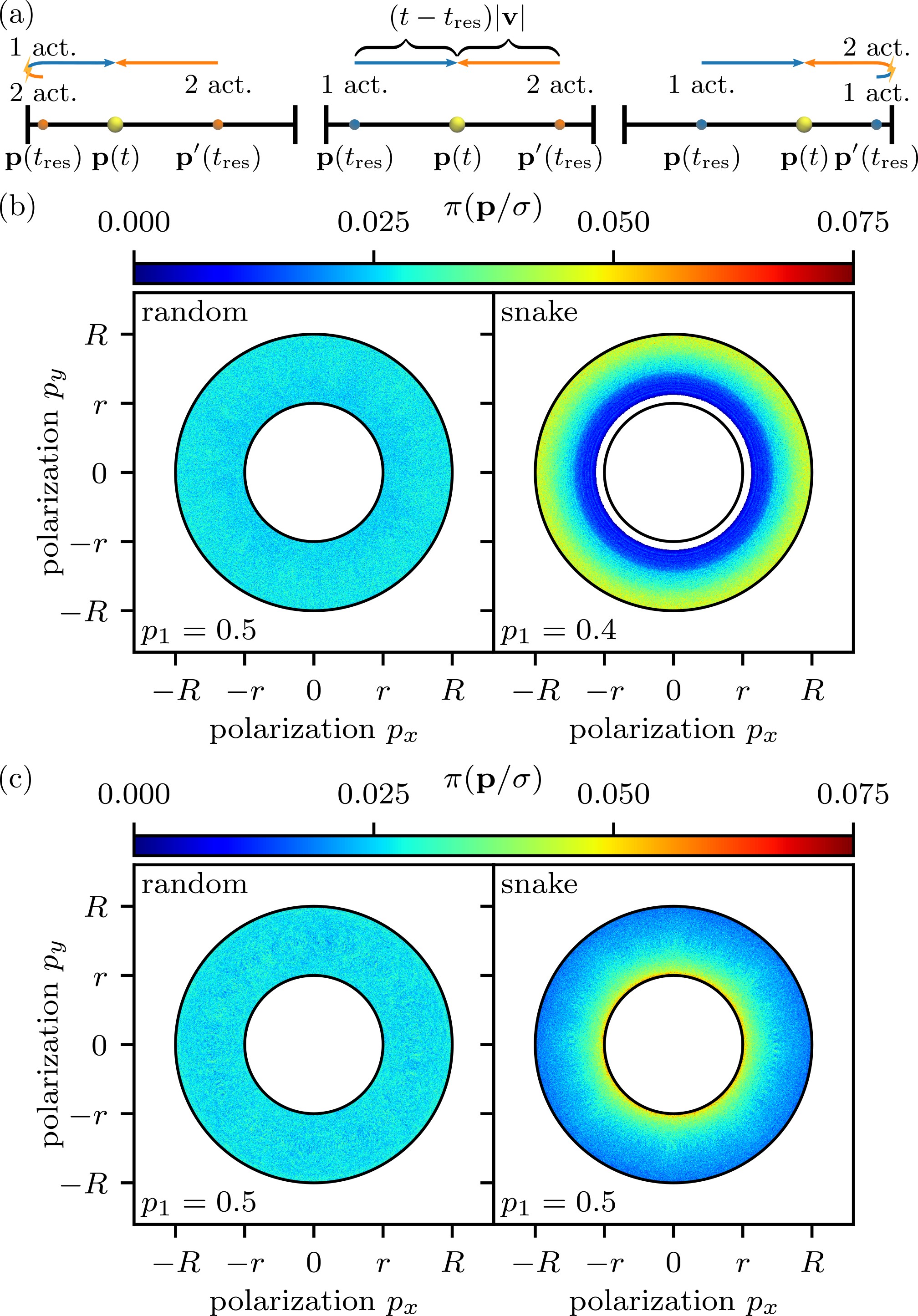}
    \caption{Global balance and irreducibility for sequential ECMC. \subcap{a} 
Flow into polarization  $\pfinal$ from $\pinitial$ and $\pinitialp$ on a 
segment of $\Omega$ as in \subfig{fig:trajectories}{a}. Events take place at 
the boundaries of the 
segment, and the active disk changes. \subcap{b} Single-dipole 
distribution  $\pi(\pvec / \sigma)$ for sequential ECMC with small 
$\Delta\phi$. The 
snake-mode distribution is incorrect, and the probability $p_1$ that disk $1$ 
is 
active right after a resampling is smaller than $p_2$. \subcap{c} Same as 
(b) but for different 
initial configuration and chain time. The snake-mode distribution is incorrect 
although both disks are equally likely to be active after a resampling ($p_1 = 
p_2$).}
    \label{fig:straight}
\end{figure}

At a resampling of straight ECMC at time $\tresamp$ (usually a multiple of 
$\tchain$), the angle $\phi$ is updated, but the active disk must also be chosen 
randomly in order to satisfy global balance. To motivate this, we show for $N=1$ 
that the two disks $1$ and $2$ must be active with the same probability at time 
$\tresamp$.  Up until the next resampling, all polarizations $\pvec(t)$ lie on a 
segment of the ring $\Omega$, and each of the  two directions  of motion on that 
segment corresponds to one of the two disks being active (see 
\subfig{fig:straight}{a}). Global balance requires the flows into $\pfinal$ to 
be independent of its position. It can be reached from two polarizations 
$\pinitial$ and $\pinitialp$. Depending on the value of $\pfinal$, the 
corresponding polarizations  $\pinitial$ and $\pinitialp$ either both correspond 
to disk $1$ being active at $\tresamp$, or both to disk $2$, or one to $1$ and 
one to $2$ (see \subfig{fig:straight}{a}). Disks $1$ and $2$ must thus be 
equally likely ($p_1 = p_2$) to be active right after the resampling, a 
\quot{random}-mode condition that is precisely implemented by the 
resampling of the active disk at $\tresamp$.

To illustrate the relevance of the random-mode requirement, we test for $N=1$ 
the case when the active disk is maintained at a resampling while the angle 
$\phi $ is 
incremented in the sequential variant of straight ECMC (\quot{snake} mode).
Clearly, snake mode is incorrect, as it samples  $\Omega$ non-uniformly
($p_1 \neq p_2$, see \subfig{fig:straight}{b}). It seems to 
be incorrect even when both disks happen to be initially active with 
practically the same probability ($p_1 = p_2$, see \subfig{fig:straight}{c}). 
In both cases, 
the snake-mode trajectory is repetitive, and lack of irreducibility in the 
lifted sample space $\Omega \times \SET{\phi}$  induces a non-uniform 
distribution in $\Omega$. For $N > 1$ at very low density, sequential 
ECMC in snake mode might not be irreducible, because a single dipole 
deterministically rotates on a closed trajectory. However, no difference
between snake mode and random mode is detected for $N>1$ at the higher 
densities considered in \sect{sec:Autocorrelations}.

\subsection{Reflective ECMC}
\label{subsec:Reflective}

Reflective ECMC~\cite{Bernard2009} differs from the straight ECMC only in its 
handling of events. At an event, the velocity of the incoming active disk is 
reflected by the line connecting the active and the target disks, and it becomes 
the velocity of the target disk. The absolute value of the velocity is thus 
preserved. For $N=1$, the polarization follows straight lines which are 
symmetrically reflected off the inner and outer ring boundaries, with a single 
reflection angle $\alpha$ (see \subfig{fig:trajectories}{b}).

Reflective ECMC satisfies global balance for any number of dipoles. For $N=1$, 
and for an initial velocity that is nearly perpendicular to the separation 
between the two disks, the polarization may follow a 
\quot{whispering-gallery-mode} trajectory which never visits small separations 
$\rho \gtrsim r$ (compare with Ref.~\onlinecite[Fig.~3]{BouchardCote2018}), 
showing that resamplings of the active disk and of its velocity angle are 
required for irreducibility. For all initial velocities at $N=1$, even for those 
that visit all of sample space $\Omega$, the probability distribution in the 
lifted sample space does not separate into independent distributions, and the 
sampled stationary distribution in $\Omega$ is non-uniform, and thus incorrect. 
With periodic resamplings, the stationary distribution is uniform in $\Omega$. 
For $N > 1 $ 
dipoles, in a large enough box, reflective  ECMC fails to be irreducible 
without resamplings, as the dynamics is then deterministic 
and a single dipole may effectively rotate in a stationary closed trajectory. 
Nevertheless, at high enough density, we have not observed any 
difference in static observables with or without resamplings.
It thus appears 
safe to omit them in practical simulations.
As we will see in  \sect{sec:Autocorrelations}, resamplings do not improve the 
decorrelation. 
Without resamplings, 
reflective ECMC is free of intrinsic parameters, and somewhat easier
to set up. Computational overhead is avoided. 

In previous work on hard disks~\cite{Bernard2009} (rather than hard-disk 
dipoles), 
reflective ECMC was benchmarked against straight ECMC and found to be 
considerably slower. In contrast, in the dipole model studied 
here, reflective ECMC will prove more powerful than straight ECMC, even when 
the latter runs with optimized intrinsic parameters.

\subsection{Forward ECMC}
\label{subsec:Forward}

Forward ECMC~\cite{Michel2020} resembles reflective ECMC in that, after an 
event, the target-disk velocities of both algorithms are located in the same 
quadrant of the coordinate system with axes parallel and orthogonal to the line 
connecting the disks at the event. All velocities are of unit absolute value, 
but forward ECMC incorporates a random element into each event. More precisely, 
the component orthogonal to the line connecting the disks at contact is 
uniformly sampled between $0$ and $1$ (while reflecting the orthogonal 
orientation). 
Its parallel component is determined so that the velocity is of unit norm. For a 
single dipole, the  polarization describes straight lines that are reflected off 
the inner and outer ring boundaries. However, the outgoing reflection angle is 
non-deterministic (see \subfig{fig:trajectories}{c}). Forward ECMC is 
irreducible  even for $N=1$, and it requires no resamplings. Also, as we will 
show in 
\sect{sec:Autocorrelations}, resamplings of the velocity and the active disk in 
intervals of the chain time do not speed up the 
algorithm.

\subsection{Newtonian ECMC}
\label{subsec:Newtonian}
Newtonian ECMC~\cite{Klement2019} mimics event-driven 
molecular dynamics.~\cite{Alder1957} All disks have velocity labels in addition 
to their positions. The label of  the active disk indicates  the time 
derivative 
of its position, that is, its displacement in time. All other disks are 
stationary. Events (including those where the maximum dipole 
separation is reached) take place as in molecular dynamics, with all velocity 
labels treated on an equal footing, and with equal 
masses for all disks. The 
identities of the active and the target disks are interchanged in the event. 
During the simulation,
the absolute value of the active-disk velocity varies. The equilibrium 
distribution of all velocity labels is uniform on the $2N$-dimensional unit 
sphere, and the root-mean-square velocity $v_\text{rms}$ is conserved.
The event rate per unit 
distance of the Newtonian ECMC equals the one of the other variants,
but the 
event rate per unit time is smaller by $\sqrt{\pi}/2$, because of the difference 
between $v_\text{rms}$ and the mean absolute 
value 
for the 
two-dimensional Gaussian distribution of velocities. All the 
velocity labels are used as lifting variables. At a 
resampling, these labels must be sampled from the exact equilibrium 
distribution (the 
rescaled Maxwell--Boltzmann distribution for all disks). Without resampling, 
the choice of the initial velocity labels is arbitrary. 

For $N=1$, the polarization trajectory of Newtonian ECMC  reflects off 
the inner and outer ring boundaries but considers both velocity labels at 
each event (see \subfig{fig:trajectories}{d}). Without resamplings, Newtonian 
ECMC breaks irreducibility for this single dipole, and it samples an incorrect 
stationary probability distribution.  For $N \ge 2$, we find no influence 
of observable means on the resampling rate, so that the irreducibility problem 
is probably again due to the high symmetry of the ring geometry in 
\sect{fig:trajectories}. In the larger-$N$ dipole systems in 
\sect{sec:Autocorrelations}, 
we furthermore notice that the algorithm becomes faster in the limit of 
infinite resampling time. 

\section{Autocorrelations for $N$ dipoles}
\label{sec:Autocorrelations}

In the present section, we characterize the local MCMC 
algorithms of \sect{sec:ECMCDescription} via the autocorrelation dynamics of the 
polarization. All correlation times $\tint$ refer to averaged integrated 
autocorrelation times of single components of the  polarization.  The unit of 
time  corresponds to a single trial move for the Metropolis algorithm or to the 
mean event time for ECMC.
In ECMC, the computational 
effort is \bigObs{1} per event, as it is per  move in the Metropolis algorithm. 
For this reason, no effort is made to compare actual CPU times, which would 
depend 
on the implementation.
Statistical errors are estimated from the 
difference between autocorrelation times of the $x$ and $y$ components of the 
polarization which, by symmetry, must be the same. For ECMC, we also take into 
account 
the uncertainty in the  mean event time.
Single  MCMC-run times, for each parameter set, are at least three orders 
of magnitude longer than $\tint$.

\subsection{Intrinsic parameters of Metropolis and straight ECMC}
\label{subsec:parameter}

\begin{figure}
\centering
\includegraphics{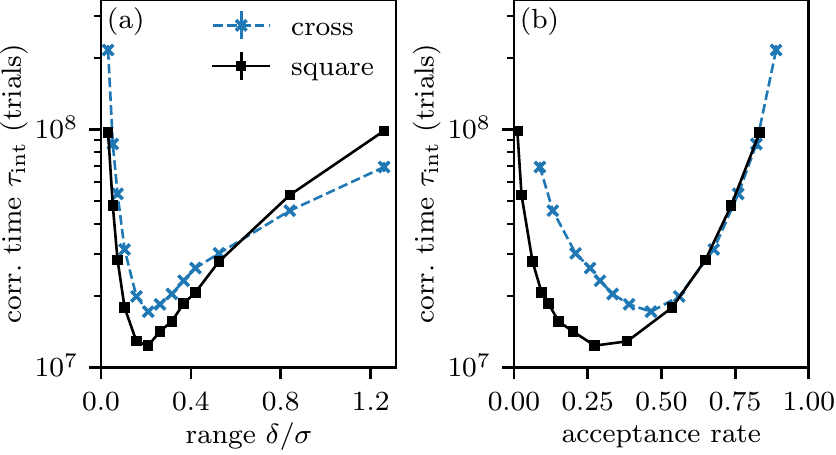}
\caption{Correlation time $\tint$ of the Metropolis 
algorithm with cross- and square-shaped displacement sets 
for $N=81$ dipoles at density $\dens=\num{0.70}$. \subcap{a} 
$\tint$ \vs\ the range $\delta$. \subcap{b} $\tint$ 
\vs\ acceptance rate. The minimum of $\tint$ provides the reference in 
\fig{fig:Comparison}, that the ECMC algorithms are benchmarked against.  
}
\label{fig:Metropolis}
\end{figure}

The correlation time of the local Metropolis algorithm depends on its intrinsic 
parameters, namely the range $\delta$ and the displacement set (see 
\subfig{fig:Metropolis}{a}). The acceptance rate decreases with increasing 
$\delta$. The correlation time can thus be expressed as a function of the 
acceptance rate. As in many comparable models, the \quot{one-half} 
rule~\cite{SMAC} is roughly respected, and the Metropolis algorithm converges 
best for a rejection rate on the order of $50\% $. For $N=81$ dipoles at 
density $\dens = 0.70$, we observe a broad optimum between 
\SIrange{20}{40}{\percent} for the square-shaped displacement set and an 
optimum 
close to $50\%$ for the cross-shaped displacement set (see 
\subfig{fig:Metropolis}{b}).

\begin{figure}
        \centering
        \includegraphics{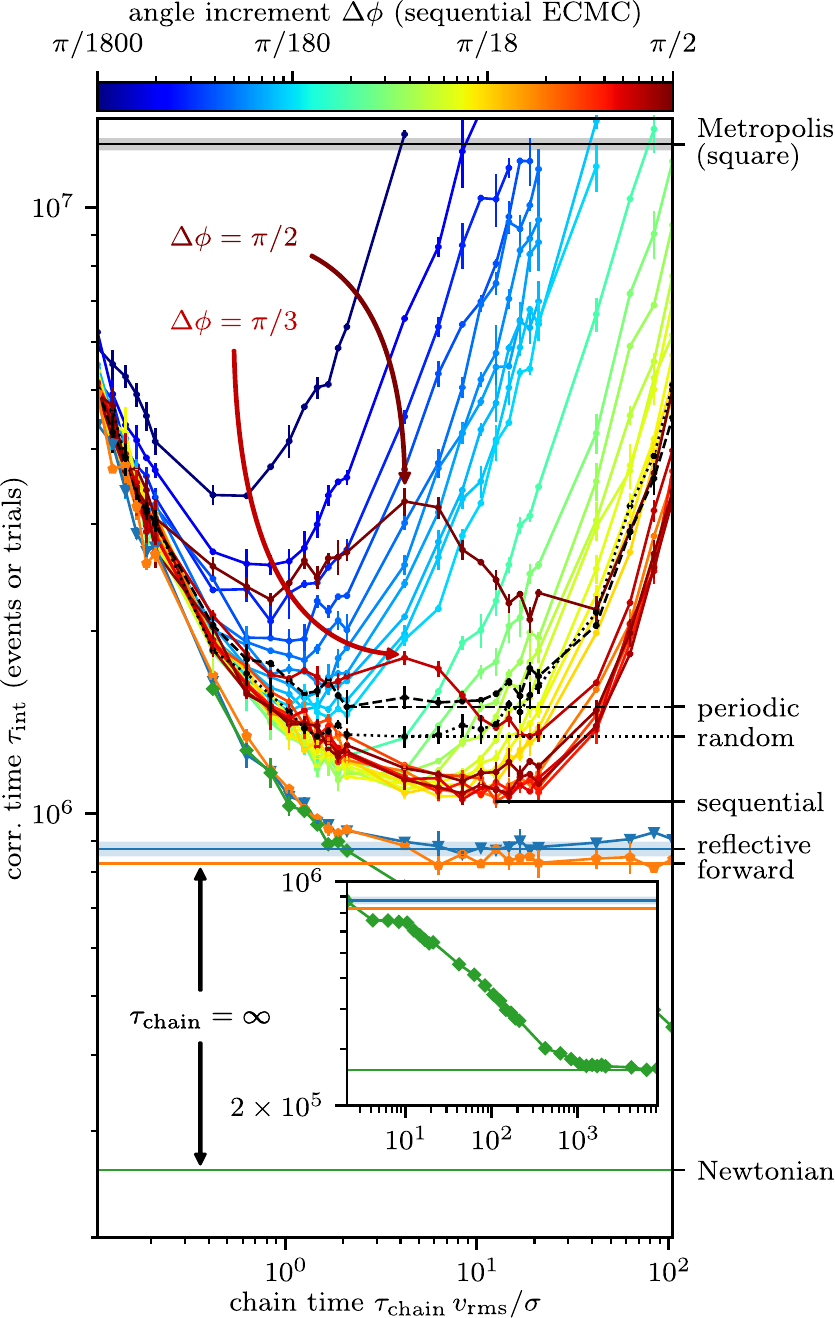}
        \caption{Correlation times of ECMC variants for different intrinsic 
parameters (chain time $\tchain$, angle increment $\Delta \phi$) for $N=81$ 
dipoles at 
density $\dens=\num{0.70}$. Optimal values of $\tint$ are highlighted on the 
right $y$-axis. For reflective, forward, and Newtonian ECMC, they agree with 
those for $\tchain \to \infty$. The optimal performance of the Metropolis 
algorithm is indicated as the benchmark reference (see \fig{fig:Metropolis}).
The inset illustrates the evolution of $\tint$ for large $\tchain$
for Newtonian ECMC. }
        \label{fig:Comparison}
\end{figure}

All ECMC variants allow for resamplings so that their correlation times
depend on the intrinsic parameter $\tchain$. The performance of the 
sequential variant of straight ECMC also depends on the angle increment $\Delta 
\phi$. Without resamplings, straight ECMC  is not irreducible and its 
correlation time is infinite. The optimum $\tint$ is thus at a finite 
$\tchain$, to be 
obtained by fine-tuning
(see upper curves in \fig{fig:Comparison}). In its
sequential variant, the cases  $\Delta \phi = \pi / 
2$ and $\Delta \phi = \pi / 3$ make the velocity cycle through a few 
values only. For these cases, we notice that $\tint$ has two local minima, 
both of which are  rather large.
In the absence of a  heuristic for the choice of intrinsic parameters, 
straight ECMC requires explicit fine-tuning, which 
increases its computational complexity.

As discussed in \sect{sec:ECMCDescription}, reflective, forward and Newtonian ECMC
appear irreducible for $N>1$ even without resampling and are fastest in this 
limit $\tchain \to \infty$ (see lower curves in \fig{fig:Comparison}). This was 
also observed, \eg, for reflective ECMC in two-dimensional hard-disk 
systems,~\cite{Bernard2009} and for Newtonian ECMC in three-dimensional hard 
spheres.~\cite{Klement2019}  For moderate values of 
$\tchain$, Newtonian ECMC is comparable to other variants. However, for much 
larger $\tchain$, its correlation times again decrease strongly (see inset of 
\fig{fig:Comparison}). For $N=81$ dipoles at density $\dens = 0.70$, the 
variants  that require no fine-tuning perform best. Newtonian ECMC without 
resampling accelerates with 
respect to the local Metropolis algorithm by a surprising factor $50$ at density 
$0.70$, that seems to  further increase at even higher density (see 
\sect{subsect:Density}).

\subsection{Density and size dependence of speedups}
\label{subsect:Density}

\begin{figure}
\centering
\includegraphics{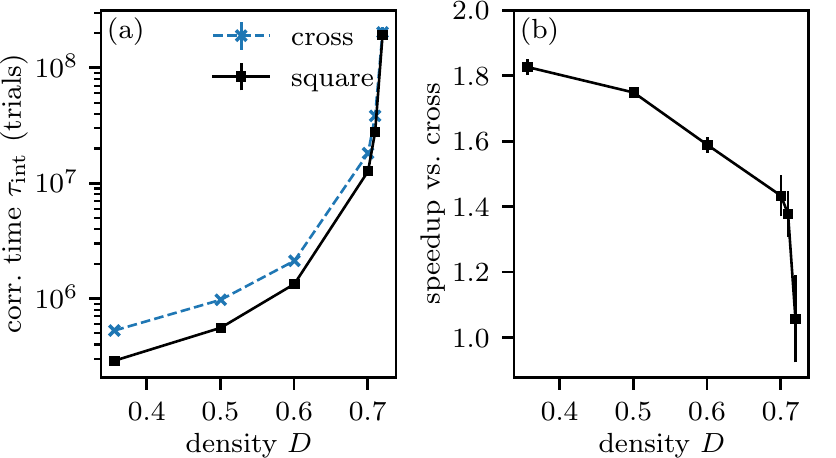}
\caption{Performance of the Metropolis algorithm for $N = 81$ dipoles
(optimized intrinsic parameters).
\subcap{a} Correlation time $\tint$ \vs\ density $\dens$ for the cross- and 
square-shaped displacement sets.
\subcap{b} Speedup for the square-shaped displacement set compared to the 
cross-shaped displacement set. }
\label{fig:MetropolisSystem}
\end{figure}

For both versions of the Metropolis algorithms, $\tint$ naturally increases at 
higher densities and climbs steeply for $\dens \gtrsim 0.70$, around the 
liquid--hexatic phase-transition density
of the hard-disk system without the tether constraint
(see 
\subfig{fig:MetropolisSystem}{a}, again for $N=81$).
For the Metropolis algorithm, the square-shaped displacement 
set is somewhat preferable over the cross-shaped displacement set for small 
densities, but the two become equally fast at large densities (see 
\subfig{fig:MetropolisSystem}{b}). This illustrates that, while step-size 
control is a key feature in reversible MCMC, the details of a local reversible 
algorithm do not really depend on the displacement sets. For simplicity, we 
chose the optimal acceptance rates for $\dens = 0.70$ for these runs at other 
densities (see \fig{fig:Metropolis}). 

\begin{figure}
	\centering
	\includegraphics{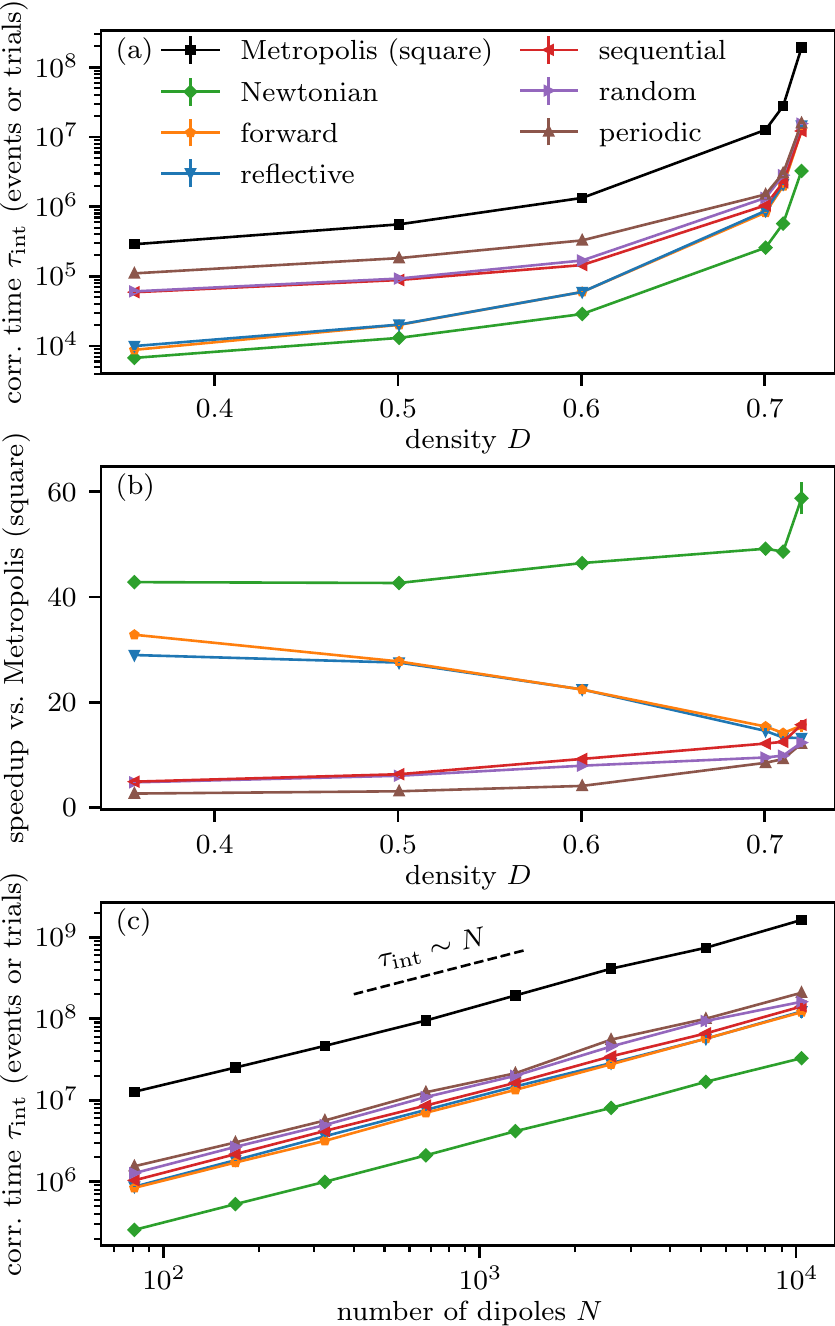}
	\caption{Performance of ECMC variants and of the Metropolis algorithm with 
		a square-shaped displacement set (where applicable: with optimized 
		intrinsic parameters). 
		\subcap{a} Correlation time $\tint$ for $N=81$ dipoles \vs\ density 
		$\dens$. 
		\subcap{b} Speedup of  ECMC with respect to the Metropolis algorithm. 
		\subcap{c} Correlation time $\tint$ at density $\dens=\num{0.70}$ \vs\ 
		number of dipoles $N$.}
	\label{fig:ECMCSystem}
\end{figure}

For the correlation times of the straight ECMC 
variants, we perform systematic scans as in \sect{subsec:parameter}, for each 
density at $N= 81$. For forward, reflective, and Newtonian ECMC, we use 
$\tchain=\infty$ without any fine-tuning.
For all ECMC variants, $\tint$ increases with 
density (see \subfig{fig:ECMCSystem}{a}). The periodic variant, 
which is fastest
for  two-dimensional hard disks,  
is the slowest variant for hard-disk dipoles. The 
random and sequential 
variants are somewhat faster. 
Reflective and 
forward 
ECMC resemble each other in performance at all $\dens$. While they are 
considerably faster than the straight variants at low $\dens$, this gap 
vanishes at $\dens=\num{0.72}$.
Newtonian ECMC is  the fastest algorithm for all  densities and, as 
mentioned, it 
requires no fine-tuning.
All ECMC algorithms outperform the Metropolis algorithm by a considerable 
margin (see 
\subfig{fig:ECMCSystem}{b}). With increasing density,
the speedup of reflective and of forward
ECMC drops from roughly $30$ to $15$, while the speedup of the straight ECMC 
variants increases with density until they become comparable. These trends were 
similarly observed near 
the liquid--hexatic transition of two-dimensional hard disks for periodic and 
reflective ECMC~\cite{Bernard2009} (where the periodic variant, however, was 
considerably faster than reflective ECMC at high densities). The 
speedup realized by Newtonian ECMC 
improves slightly from low to high densities, as was likewise found for 
three-dimensional hard spheres.~\cite{Klement2019} We observe the highest 
speedup of roughly a factor of $60$ at the highest density $\dens=\num{0.72}$ 
that we studied.

We finally study the dependence of the correlation time $\tint$ of the number of 
dipoles $N$ at the density $\dens=\num{0.70}$. For the reflective, forward and 
Newtonian ECMC variants, we simply set $\tchain = \infty$, given our findings at 
$N=81$. For periodic, random and sequential ECMC, we infer the optimal $\tchain$ 
and $\Delta\phi$ from the $N=81$ case (performing occasional sweeps through 
intrinsic parameters as cross-checks). For the Metropolis algorithm, we used 
the optimal acceptance rate of $N=81$. We observe that $\tint$ increases 
linearly with $N$ for all algorithms (see \subfig{fig:ECMCSystem}{c}). This 
implies that the observed speedups of the non-reversible algorithms stay 
constant. 

\section{Conclusions and outlook}
\label{sec:Conclusions}

In this paper, we have systematically studied  local MCMC algorithms for complex 
molecules that we caricatured as hard-disk dipoles with parameters inspired by 
the  SPC/Fw water system. Moving from three to two spatial dimensions and from a 
liquid of charged molecules to a model of hard-disk dipoles has greatly 
simplified the algorithm development and its implementation. This allowed us to 
scan thousands of parameters sets for different ECMC variants and to benchmark
them against the reversible local Metropolis algorithm. At this exploratory 
stage, this would have required  prohibitive computing resources for the 
original three-dimensional model.~\cite{Qin2020Thesis} The broad spread of  the 
performance of different ECMC versions came as a surprise. We found that 
algorithms with intrinsic parameters perform considerably less well than 
others and demonstrated 
a 60-fold speedup of Newtonian ECMC with respect to the Metropolis 
algorithm. Our study has also brought out subtleties of ECMC, that are hidden in 
simple liquids. For the case of a single dipole, we discovered strict 
requirements for resampling in order to ensure irreducibility of straight, 
reflective and Newtonian ECMC. However,  
these requirements did not seem to play a role for denser systems of dipoles. 
Event-based randomness as in forward ECMC (or as in all ECMC simulations of 
soft interactions) strictly ensures irreducibility.

We believe that our observations will carry over from the caricature 
two-dimensional dipoles to the three-dimensional SPC/Fw systems and related 
models. There are two reasons why we have not yet studied this system directly. 
One is, as mentioned, the scale of the computing requirements for a 
full-fledged three-dimensional scan. The 
other is that further 
algorithm development is still required. The cell-veto 
algorithm~\cite{KapferKrauth2016} (which reduces the computational complexity 
of the Coulomb problem to \bigObs{N \log N} per sweep of events, without 
evaluating the Coulomb energy) is in the present version of our open-source 
project  only implemented for the periodic variant of straight ECMC. However, we 
expect it to generalize  to  all ECMC variants for flexible water 
molecules with an explicit Coulomb interaction. The benchmark against Metropolis 
MCMC will then be much more favorable, as the change of the Coulomb energy
can there only
be computed in \bigObs{ N^{3/2}} per sweep of 
moves.~\cite{BerthoumieuxMaggs2010} The key question will be whether the dynamic 
arrest of straight ECMC for the three-dimensional water system can be overcome 
as in the two-dimensional hard-disk dipole model studied here. The extreme 
dependence of the performance of non-reversible MCMC on details of the algorithm 
was also evidenced in the escape times from a tightly confined initial 
configuration,~\cite{Hoellmer2021} which might be relevant for overcoming 
dynamic arrest. 

Finally, we point out that any tethered-dipole configuration is also a valid 
hard-disk configuration, with an added upper limit  on the distance between 
pairs of hard disks. For the tethered dipole system, the global hexatic 
orientational order parameter $\psi_6$ (see 
Refs.~\onlinecite{Nelson1979,Bernard2011}), at densities $\dens \gtrsim 0.70$, 
decorrelates faster than for the  hard-disk system which there already enters 
the mixed hexatic--fluid phase. This observed effect might allow for an 
improved sampling algorithm for hard disks. It is also unclear why straight 
ECMC performs worse for dipoles than the reflective, forward, and Newtonian 
variants, while for simple hard disks straight ECMC is clearly the most 
efficient.

\acknowledgments

P.~H.~acknowledges support from the Studienstiftung des deutschen Volkes and 
from Institut Philippe Meyer. 
W.~K.~acknowledges support from the Alexander von Humboldt Foundation.

\section*{Additional remarks}
The figures in this paper were prepared with the Matplotlib~\cite{matplotlib} 
and Visual Molecular Dynamics~\cite{VMD} (VMD) software packages.
Integrated autocorrelation times were computed by 
the Python \texttt{emcee} package.~\cite{emcee}

\bibliography{General,DipoleMany2020}

\end{document}

%% file: latexcommands_2021d.tex
\newcommand\subcap[1]{{(#1):}}

\newcommand{\SET}[1]{\{#1\}}

\newcommand{\fig}[1]{Fig.~\ref{#1}}

\newcommand{\quot}[1]{``#1''}

\newcommand{\sect}[1]{Section~\ref{#1}}

\newcommand{\eg}{\textrm{e.g.}}

\newcommand{\vs}{\textrm{vs.}}

\newcommand{\FCAL}{\mathcal{F}}  
\newcommand{\OCAL}{\mathcal{O}}  

\newcommand{\VEC}[1]{\mathbf{#1}}
\newcommand{\pvec}{\VEC{p}}

\newcommand\bigObs[1]{\ensuremath{\OCAL  (#1)}}